\documentclass[letterpaper,titlepage,11pt]{article}
\usepackage{hyperref}
\usepackage{amssymb,amsmath,amsfonts}
\usepackage{epsfig}
\usepackage{graphicx}
\usepackage{float}
\usepackage[english]{babel}
\usepackage[T1]{fontenc}
\usepackage[applemac]{inputenc}

\def\clock{{\count0=\time
           \divide\count0 60
           \ifnum\count0<10 0\fi\the\count0
           \multiply\count0 -60 \advance\count0 \time
           :\ifnum\count0<10 0\fi \the\count0
         }}
\newcommand{\timestamp}{{\small\vbox{\hbox{\tt\jobname.tex}
\hbox{\the\day/\the\month/\the\year, \clock}}}}

\newcommand{\Z}{\mathbb{Z}}

\newcommand{\R}{\mathbb{R}}

\newcommand{\ds}{\displaystyle}

\numberwithin{equation}{section}

\begin{document}

\begin{titlepage}

\begin{flushright}
NORDITA-2009-74\\
\end{flushright}

\vskip 2cm

\centerline{\huge Uniqueness Theorem for Black Hole Space-Times }
\vskip 0.2cm \centerline{\huge with Multiple Disconnected Horizons}

\vskip 1.6cm
\centerline{\bf  Jay Armas$^a$\footnote{jay@nbi.dk} and Troels Harmark$^b$\footnote{harmark@nordita.org}}
\vskip 1.1cm
\centerline{\sl $^a$The Niels Bohr Institute}
\centerline{\sl Blegdamsvej 17, DK-2100 Copenhagen \O, Denmark}
\vskip 0.5cm
\centerline{\sl $^b$NORDITA}
\centerline{\sl Roslagstullsbacken 23,
SE-106 91 Stockholm,
Sweden}

\vskip 2cm

\centerline{\bf Abstract} \vskip 0.2cm \noindent We show uniqueness
of stationary and asymptotically flat black hole
space-times with multiple disconnected horizons and with two
rotational Killing vector fields in the context of five-dimensional
minimal supergravity (Einstein-Maxwell-Chern-Simons gravity). The
novelty in this work is the introduction in the uniqueness theorem
of intrinsic local charges measured near each horizon as well as the measurement of local fluxes besides the asymptotic charges that characterize a particular solution. A
systematic method of defining the boundary conditions on the fields
that specify a black hole space-time is given based on the study of
its rod structure (domain structure). Also, an analysis of known
solutions with disconnected horizons is carried out as an example of
an application of this theorem.

\vskip 3cm

\begin{flushright}
\emph{"But the perfect scientist is also a gardener: he believes that beauty is knowledge."} \\
 Gon\c calo M. Tavares in \underline{Brief Notes on Science}
\end{flushright}


\end{titlepage}

\small
\tableofcontents
\normalsize
\setcounter{page}{1}


\section{Introduction and Summary}

Since the discovery of the black ring \cite{Emparan:2001wn} it has
become increasingly clear that the subject of higher-dimensional
black holes is very rich and interesting. With the finding of the
five-dimensional black ring solution it became clear that unlike in
four dimensions black holes are not unique when given the
asymptotically measurable conserved quantities. In fact, even
restricting to a particular topology of the event horizon - namely
that of the black ring - there can be two regular black ring phases
available given the asymptotically measurable conserved quantities,
$i.e.$ the mass and angular momenta for pure gravity solutions.

The pressing question is thus whether there exists some meaningful
generalization of the uniqueness properties of four-dimensional
black holes to higher dimensions. This is important in order to
understand the "space of black holes", $e.g.$ how many black holes
there are and how to classify them. It would also be crucial if one
is given two strongly curved space-times with event horizons and one
is asked to determine whether they are different or not. One would
therefore like a finite list of invariants that can fully
characterize a black hole space-time. The list of invariants include
topological invariants (such as the event horizon topology),
geometrical invariants ($e.g.$ lengths, areas, volumes etc.) and
locally and globally measured physical quantities such as the mass,
angular momenta, charges, fluxes and so on.

The rod structure introduced in \cite{Harmark:2004rm} can be used to
characterize and classify asymptotically flat stationary black holes
in five-dimensional pure gravity, assuming the existence of two
commuting rotational Killing vector fields corresponding to the
rotations in two orthogonal planes. Indeed, it was proven in
\cite{Morisawa:2004tc,Hollands:2007aj}%
\footnote{See also \cite{Rogatko:2008yd}.} that for solutions with a
single horizon you can uniquely characterize a solution given the
asymptotic charges and the rod structure. Specifically, this means
that if you in addition to the asymptotic charges also specify the
topology of the event horizon and the lengths of the rods, you have
a full characterization. Thus, the two possible regular black ring
phases can be distinguished by the lengths of the rods. Therefore,
for single horizon, asymptotically flat, stationary black hole
space-times in five-dimensional pure gravity we can make a list of
invariants that includes the topological invariants given by the
rod-directions, the geometrical invariants given by the lengths of
the rods and finally the asymptotically measured mass and angular
momenta.

Subsequently several works expanding these results have been found.
This includes \cite{Hollands:2007qf} where the uniqueness theorem
for a single horizon asymptotically flat black hole in
five-dimensions was generalized to Einstein-Maxwell (EM) gravity in
five dimensions, \cite{Hollands:2008fm} where the interesting
structure of invariants for black holes with Kaluza-Klein space
asymptotic were examined and the uniqueness theorem of
\cite{Hollands:2007aj} was developed further, \cite{Tomizawa:2009ua}
where the uniqueness of charged rotating black holes with spherical
topology in five-dimensional Einstein-Maxwell-Chern-Simons (EMCS)
gravity was proven  and \cite{Amsel:2009et,Figueras:2009ci} where the
uniqueness of extremal black holes were considered.%
\footnote{While finishing up the first preprint version of this paper the
paper \cite{Tomizawa:2009tb} appeared as preprint with
a uniqueness theorem for charged dipole black rings in five-dimensional EMCS gravity which therefore overlap with our independent results on this. As stated below, we have used some results of \cite{Tomizawa:2009tb} in this version of the paper to further enhance our results.} Finally, in
\cite{Harmark:2009dh} the concept of rod-structure, which also has
been developed further in
\cite{Hollands:2007aj,Hollands:2007qf,Hollands:2008fm}, was
significantly generalized to include black hole space-times with any
type of matter fields and any number of Killing vector fields. The
generalized rod-structure is named {\sl domain-structure}.%
\footnote{This is because when one has less than $D-2$ Killing
vector fields in the black hole space-time the boundary of the orbit
space is no longer one-dimensional thus neither the name
rod-structure nor interval-structure is suitable. Moreover, the
domain-structure of \cite{Harmark:2009dh} can be defined without
using the Einstein equations, unlike in
\cite{Harmark:2004rm,Hollands:2007aj,Hollands:2007qf,Hollands:2008fm},
hence in general it is not natural to think of each domain as
associated with a source.} The domain-structure can be further
generalized to non-asymptotically flat black hole space-times
\cite{Harmark:2010ad}.

In this paper we consider asymptotically flat stationary black holes
with two commuting rotational Killing vector fields in
five-dimensional EMCS gravity. We generalize the existing uniqueness
theorems for five-dimensional asymptotically flat black holes in two
respects. First, we allow for the first time for disconnected event
horizons, $i.e.$ we allow for arbitrarily many distinct connected
event horizons. This is an important generalization in that it
allows one to see which invariants should be defined locally with
respect to each black hole, or each domain of the space-time, and
which invariants that should be measured asymptotically. In the
special case where one consider black holes without the gauge field
turned on the uniqueness theorem generalizes that of
\cite{Hollands:2007aj}. Secondly, we generalize the uniqueness
theorem of \cite{Tomizawa:2009ua} for single horizon charged
spherical black holes in five-dimensional EMCS gravity to include
the horizon topology of the black ring and we consider therefore
both the charge of each event horizon and the two kinds of dipole
charges that one has, corresponding to the two possible orientations
of black rings. To prove the uniqueness theorem we employ the
techniques of \cite{Tomizawa:2009ua} based on the sigma-model
construction of \cite{Bouchareb:2007ax}.

We choose in this paper to consider asymptotically flat stationary
black hole space-times in five-dimensional EMCS gravity. This is
motivated by the fact that EMCS gravity is the bosonic part of the
five-dimensional minimal supergravity theory, which again is a low
energy limit of string theory. This five-dimensional minimal
supergravity shows up in many interesting string theory related
contexts and it is important to obtain a characterization of black
hole space-times in EMCS gravity. Another reason that we consider
EMCS gravity is that one can find a sigma-model description with a
$G_{2(2)}$ symmetry \cite{Bouchareb:2007ax,Compere:2009zh}. This
makes it possible to make powerful solution generating techniques
based on the integrability of the Einstein equations of EMCS
gravity, similarly to what has been done for the five-dimensional
pure gravity case \cite{Pomeransky:2005sj}. Finally, many
interesting black hole solutions in EMCS gravity have been found
\cite{Emparan:2004wy,Elvang:2007rd,Yazadjiev:2007cd,Evslin:2007fv,Iguchi:2007bd,Yazadjiev:2008pt,Elvang:2007hs}
which makes it important to study their uniqueness properties.

We emphasize that we in particular are able to handle the non-trivial situation where there are fluxes in between the event horizons of the multi-black hole solution. In fact, the relevance of the fluxes to characterize asymptotically Kaluza-Klein black hole solutions has been pointed out in \cite{Yazdjiev:2009mgb}. These fluxes are measured as the fluxes through the minimal surfaces in between black rings or in between a black ring and a Myers-Perry black hole. There are two types of fluxes, one being the standard magnetic flux measured from the two-form field strength $F=dA$ while the other flux is measured from the Chern-Simons contribution $A \wedge F$ that also appears in the definition of charge. For non-zero fluxes and dipole charges the uniqueness theorem involves boundary conditions with arbitrary functions. However, as first found in \cite{Tomizawa:2009tb} for the dipole black ring and black lens, the uniqueness theorem can work despite this fact.

A further motivation to this work came from the dipole ring solution
\cite{Emparan:2004wy} which is a solution of both five-dimensional
EM and EMCS gravity. Since the dipole charge of the black ring is
defined from a contractible circle it is not a conserved quantity.
This results in a continuous non-uniqueness of black ring phases
when given the asymptotic charges, $i.e.$ the mass, angular momenta
and the total charge. In this paper we prove that a dipole black ring in
EMCS gravity is unique given its asymptotic charges, rod structure
and dipole charge. Thus, as expected, the dipole charge provide a
further local degree of freedom for black hole solutions. Note that
for the case of EM gravity the dipole charge was included in the
uniqueness theorem of \cite{Hollands:2007aj} though with certain
constraints on the gauge fields.%
\footnote{While finishing up the first version of this paper the
paper \cite{Tomizawa:2009tb} appeared as preprint with
a uniqueness theorem for dipole black rings in five-dimensional EMCS gravity.}

We add two extra ingredients to our uniqueness theorem as well: The
general definition of rod structure in EMCS gravity and the general
definition of dipole charge in terms of the rod structure. The
definition of the rod structure we can infer from the completely
general definition of domain structure of \cite{Harmark:2009dh}
employed in the special case of five-dimensional EMCS gravity. We
describe how this works in the paper and employ the definition of
rod structure in EMCS gravity to the case of the dipole black ring
of \cite{Emparan:2004wy}. We find here a natural definition of the
two types of dipole charges in EMCS gravity based on the potentials
used in the $G_{2(2)}$ sigma-model construction of
\cite{Bouchareb:2007ax}.

Finally, as an extra bonus, we consider furthermore the uniqueness theorem for the black lens, first considered in EMCS gravity in \cite{Tomizawa:2009tb}. We consider in particular the role of the fluxes in this uniqueness theorem.%
\footnote{The uniqueness theorem for the black lens was not included in the first preprint version of this paper.}

This paper is build up as follows. In Section \ref{domain} we define
the rod-structure of black hole space-times in five-dimensional EMCS
gravity by employing the general definition of the domain structure
in \cite{Harmark:2009dh}. In Section \ref{super} we write down the
action of Einstein-Maxwell-Chern-Simons Gravity and give all the
necessary formalism needed for our uniqueness theorem, specifically,
we use the sigma model approach to show that a particular black hole
solution is characterized by a certain number of potentials. In
section \ref{uniqueness} we specify the rod structure of the most
general black hole solution with multiple disconnected horizons
which can be found in this theory and satisfies our requirements. We
then proceed by providing a systematic method to impose correct
boundary on the potentials mentioned above leading to a proof of
uniqueness for such space-times following an analysis of the
uniqueness of different known exact solutions is taken. Further, we show how to generalize the theorem to include black space-times with Lens space horizon topology. In section
\ref{discussion} we discuss the implications and limitations of our
results.

\section{From Domain Structure to Rod Structure }\label{domain}

We consider a five-dimensional asymptotically flat stationary black
hole space-time. We assume that it has three commuting Killing
vector fields $V_{(0)}$, $V_{(1)}$ and $V_{(2)}$ such that $V_{(0)}$
is asymptotically time-like and generates $\R$ while $V_{(1)}$ and
$V_{(2)}$ both are space-like and each generates a $U(1)$. In
\cite{Harmark:2009dh} it is shown that the metric of the black hole
space-time can be put in the form
\begin{equation} \label{E:dscan}
\begin{array}{c} \ds
ds^2 = G_{ij}(dx^{i}+A^{i})(dx^{j}+A^{j}) + e^{2\nu} ( dr^2 +
\lambda^2 dz^2 )
\\[2mm]
r^{2}=|\det G_{ij}|,~ \lambda\to1 ~ \mbox{for} ~ r\to\infty
\end{array}
\end{equation}
where $i,j=1,2,3$, $G_{ij}$ and $z \in \R$. This is shown without
use of the Einstein equations and it is explained in
\cite{Harmark:2009dh} that one can use this to define the
rod-structure/domain-structure for any black hole space-time given
the above assumptions.

The set of space-like Killing vector fields $V_{(1)},V_{(2)}$
corresponds to a particular choice of basis. If we consider a new
basis $T_{(1)},T_{(2)}$ then in general it is in a linear
combination:
\begin{equation} \label{E:kvb}
T_{(i)}=\sum^{2}_{j=1}U_{ij}V_{(j)}
\end{equation}
We want each of the $T_{(i)}$ to generate a $U(1)$ isometry and we
choose the period of the flow of the Killing vector fields to be
$2\pi$. This together with the fact that the above transformation
should be invertible comes the requirement that
$U~\in~GL(p-1,\mathbb{Z})$ with $\det(U)=\pm1$. Hence we are not
entirely free to choose the basis.

Requiring the following two conditions to hold:
\begin{itemize}
\item[(i)] $V_{(0)}^{[\mu_{1}}V_{(1)}^{\mu_{2}} V_{(2)}^{\mu_{3}}D^{\nu}V_{(i)}^{\rho]}=0$ for at least one point of space-time for a given $i=0,1,2$.
\item[(ii)] $V_{(i)}^{\nu}R_{\nu}^{[\rho}V_{(0)}^{\mu_{1}}V_{(1)}^{\mu_{2}} V_{(2)}^{\mu_{3}]}=0$ for all $i=0,1,2$.
\end{itemize}
then the two-planes orthogonal to the Killing vector fields
$V_{(i)}$ are integrable, and we are free to set $A^i=0$ and
$\lambda=1$ everywhere. The metric \eqref{E:dscan} is then reduced
to the simpler form:
\begin{equation} \label{E:dsred}
ds^{2}=G_{ij}dx^{i}dx^{j} + e^{2\nu}(dr^{2}+dz^{2}), ~ r^{2}=|\det
G_{ij}|
\end{equation}
In the class of black hole space-times we consider in this paper the
second condition is guarantied by the Einstein equations of
Einstein-Maxwell-Chern-Simons (EMCS) gravity (we write the action
below in Section \ref{super}), while for the first condition one can
use that we require the space-times to be asymptotically flat.

In line with \cite{Harmark:2009dh} we define now the rod-structure
(domain-structure) of the solution by analyzing the behavior of
$G_{ij}$ for $r=0$ which constitutes the $z$-axis in the coordinates
\eqref{E:dsred}. For $r=0$ we have $\det G_{ij}=0$. Let $Q_2$ be the
set of points at $r=0$ for which $\dim \ker G \geq 2$. This can be
shown to be a set of points $\kappa_i$, $i=1,2,...,p$, with
$\kappa_i < \kappa_{i+1}$, that defines a set of intervals $( -
\infty, \kappa_1)$, $(\kappa_1,\kappa_2)$, ... , $(\kappa_p,\infty)$
\cite{Harmark:2009dh}. These intervals are the rods (domains) of the
black hole space-time. On the inner part of the intervals $\dim \ker
G = 1$. The direction of each interval (rod) correspond to the
direction of $v \in \ker G$ at the interval. In case the direction
of a rod is time-like for $r \rightarrow 0$ the rod defines a
Killing horizon with the direction being that of the corresponding
Killing vector field. Otherwise the direction is space-like for
$r\rightarrow 0$ and the rod corresponds to a fixed plane of
rotation.

In case of a rod with a space-like direction $v$ the above
restriction on the change of basis of the $U(1)$ Killing vector
fields means that $v= m V_{(1)} + n V_{(2)}$ where $m,n \in \Z$.
Furthermore, two rods with space-like directions $v= m_1 V_{(1)} +
n_1 V_{(2)}$ and $v'= m_2 V_{(1)} + n_2 V_{(2)}$ obey $m_1 n_2 - m_2
n_1 = \pm 1$. Thus, we regain the restrictions on the rod-structure
obtained in \cite{Hollands:2007aj} now for the case of EMCS gravity
(see also \cite{Harmark:2009dh}).

The rod-structure (domain-structure) of the black hole space-time
thus consists of two sets of invariants
\begin{itemize}
\item A set of topological invariants: The split up of the $z$-axis
into intervals, with each rod either being a Killing horizon or a
fixed plane of a rotation. Furthermore, for the space-like rods
corresponding to a fixed plane of rotation the direction is $v= m
V_{(1)} + n V_{(2)}$ with $m,n \in \Z$ (along with the above restriction on
successive rods). This constitutes a set of topological invariants
of the black hole space-time.
\item A set of geometrical invariants: The lengths of the rods,
measured simply as $\kappa_{i+1}-\kappa_i$ for the rod
$(\kappa_i,\kappa_{i+1})$.
\end{itemize}
The above set invariants are the invariants that one can read off
from the metric of a black hole space-time in EMCS gravity. Below we
shall use the matter fields of EMCS gravity to define further
invariants in the form of locally and globally measured physical
quantities such as the mass, angular momenta, and various types of
charges and fluxes and explore what set of invariants can
characterize uniquely a rather large set of black hole space-times
in EMCS gravity.

\section{Minimal Supergravity} \label{super}

In this section we write down the necessary formalism for
five-dimensional Einstein-Maxwell-Chern-Simons (EMCS) gravity, also
known as (the bosonic sector of) minimal supergravity, which will be
useful for the uniqueness theorem of section \ref{uniqueness}. We
first give the original action of 5D EMCS gravity and then by
writing the metric in the Weyl-Papapetrou form we rewrite this
action in the non-linear sigma model form for which a Mazur identity
can be derived.

\subsection{Action and Weyl-Papapetrou form}

The action and the equations of motion are given by:

\begin{equation} \label{E:minS}
S=\frac{1}{16 \pi}[ \int dx^{5} \sqrt{-g} (R - \frac{1}{4}F^{2}) - \frac{1}{3\sqrt{3}} \int F\wedge F\wedge A]
\end{equation}

\begin{equation} \label{E:mot1}
R_{\mu\nu} - \frac{1}{2}R g_{\mu \nu}=\frac{1}{2}(F_{\mu \lambda}F_{\nu}^{\lambda} - \frac{1}{4}g_{\mu \nu}F_{\rho \sigma} F^{\rho \sigma})
\end{equation}
\begin{equation} \label{E:mot2}
d*F + \frac{1}{\sqrt{3}}F\wedge F=0
\end{equation} \\
Now, define the Killing vector fields $V_{(0)}=\frac{\partial}{\partial t},~V_{(1)}=\frac{\partial}{\partial \phi},~V_{(2)}=\frac{\partial}{\partial \psi}$ and assume that $V_{(1)}$ and $V_{(2)}$ also preserve the Maxwell field, i.e., $L_{V_{(a)}}F=0$ for $a=1,2$. Then, the metric \eqref{E:dsred} can be rewritten in the Weyl-Papapetrou form:
\begin{equation} \label{E:dswp}
ds^{2}=\lambda_{ab}(dx^{a} + a^{a}_{t}dt)(dx^{b} + a^{b}_{t}dt)  + \tau^{-1}(e^{2\sigma}(dr^{2} + dz^{2}) - r^{2}dt^{2}), ~\tau=-det(\lambda_{ab})
\end{equation}
together with the gauge field \footnote{Due to gauge freedom one can
always set $A_{r}=A_{z}=0$ (see \cite{Tomizawa:2009ua}).}:

\begin{equation} \label{E:gau}
A=\sqrt{3}\psi_{\phi}d\phi + \sqrt{3}\psi_{\psi}d\psi + A_{t}dt
\end{equation}
In Appendix \ref{aa} we give the relations between \eqref{E:dsred} and \eqref{E:dswp}.\\ \\
The metric functions $a^{a}_{t}$ are determined by:

\begin{equation} \label{E:atr}
a^{a}_{t,r}=r\tau^{-1}\lambda^{ab}(\omega_{b,z} -3 \psi_{b}\mu_{,z} - \psi_{b}\epsilon^{cd}\psi_{c}\psi_{d,z})
\end{equation}
\begin{equation} \label{E:atz}
a^{a}_{t,z}=-r\tau^{-1}\lambda^{ab}(\omega_{b,r} -3 \psi_{b}\mu_{,r} - \psi_{b}\epsilon^{cd}\psi_{c}\psi_{d,r})
\end{equation}
And the t-component of the gauge field by:

\begin{equation} \label{E:Atr}
A_{t,r}=\sqrt{3}[a^{a}_{t}\psi_{a,r} - r\tau^{-1}(\mu_{,z}+\epsilon^{bc}\psi_{b}\psi_{c,z})]
\end{equation}
\begin{equation} \label{E:Atz}
A_{t,z}=\sqrt{3}[a^{a}_{t}\psi_{a,z} + r\tau^{-1}(\mu_{,r}+\epsilon^{bc}\psi_{b}\psi_{c,r})]
\end{equation}
The electric, magnetic and twist potentials $\psi_{a},~\mu,~\omega_{a}$ are respectively determined by:

\begin{equation} \label{E:epot}
d\psi_{a}=-\frac{1}{\sqrt{3}}i_{V_{(a)}}F
\end{equation}
\begin{equation} \label{E:u}
d\mu=\frac{1}{\sqrt{3}}B - \epsilon^{ab}\psi_{a}d\psi_{b}
\end{equation}
\begin{equation} \label{E:twi}
d\omega_{a}=W_{a} + \psi_{a}(3d\mu + \epsilon^{bc}\psi_{b}d\psi_{c})
\end{equation}
with the magnetic one-form $B$ and the twist one form $W_{a}$ given by:
\begin{equation} \label{E:b}
B=*(V_{(1)}\wedge V_{(2)}\wedge F)
\end{equation}
\begin{equation} \label{E:v}
W_{a}=*(V_{(1)}\wedge V_{(2)}\wedge dV_{(a)})
\end{equation}
Here $V_{(i)}$, $i=1,2$, are the one-forms gotten from the Killing vector fields $V_{(i)}$, $i=1,2$, by using the metric.

\subsection{Reduction to the Non-Linear Sigma Model and the Mazur Identity}

With all of this equipment one can rewrite the action \eqref{E:minS}
in the non-linear sigma model form (see \cite{Tomizawa:2009ua} for
details):

\begin{equation} \label{E:nonS}
S=\frac{1}{4} \int dr dz tr(\Theta^{-1} \partial_{i}\Theta \Theta^{-1} \partial^{i}\Theta)
\end{equation}
where $\Theta$ is a $7$x$7$ matrix defined as:

\[ \Theta = \left( \begin{array}{ccc}
\hat{A} & \hat{B} & \sqrt{2}\hat{U} \\
\hat{B^{T}} & \hat{C} & \sqrt{2}\hat{V} \\
\sqrt{2}\hat{U^{T}} & \sqrt{2}\hat{V^{T}} & \hat{S} \end{array} \right)\] \\
where $\hat{A}$ and $\hat{C}$ are symmetric 3x3 matrices, $\hat{B}$ is a 3x3 matrix, $\hat{U}$ and $\hat{V}$ are 3-component column matrices, and $\hat{S}$ is a scalar. These entries depend only on $\lambda_{ab}, \phi_{a}, \mu, \omega_{a}$. Their explicit form is given in Appendix \ref{ab}. $\Theta$ has the property of being symmetric and unimodular ($det\Theta=1$) and can be split as $\Theta=\hat{g}\hat{g}^{T}$ with $\hat{g}$ being a $G_{2(2)}$ matrix.\\ \\
If we now consider two different field configurations $\Theta_{0}$ and $\Theta_{1}$, i.e., two different configurations of $\{\lambda_{ab},  \omega_{a}, \mu, \psi_{a}\}$, then one can derive the Mazur identity:

\begin{equation} \label{E:Mi}
\int_{\partial \Sigma} r \partial_{a} tr{\Psi} dS^{a}=\int_{\Sigma} r h_{ab}tr(M^{Ta} M^{b}) dr dz
\end{equation}
with $\Psi$ defined as:

\begin{equation} \label{E:Psi}
\Psi = \Theta_{1}\Theta^{-1}_{0} - 1
\end{equation}
and $h=dr^{2} + dz^{2}$, whereas, M is given by:

\begin{equation} \label{E:M}
M^{a}=\hat{g}_{0}^{-1}\bar{J}^{Ta}\hat{g}_{1}
\end{equation}\\
with $\bar{J}$ defined as $\bar{J}^{a}=J_{1}^{a} - J_{0}^{a}=\Theta^{-1}_{1} \partial^{a}\Theta_{1} - \Theta^{-1}_{0} \partial^{a}\Theta_{0}$.\\ \\
The integral over the boundary $\partial \Sigma$ in \eqref{E:Mi} is
taken over the $z$-axis at $r=0$ and at infinity. If the LHS of
\eqref{E:Mi} vanishes then we must have $\bar{J}=0$ on the RHS.
Hence, if $\bar{J}=0$ the matrix $\Psi$ must be constant over the
entire region $\Sigma$. It then suffices to show that $\Psi$ is zero
at one part of the boundary $\partial \Sigma$ in order to prove the
equivalence of the two solutions. This will be the basis for our
uniqueness theorem of section \ref{uniqueness} below.

\section{Uniqueness of Black Holes with Disconnected Horizons} \label{uniqueness}

Our goal in this section is to prove the following theorem: \\ \\
\textbf{Theorem 1:}  \emph{Consider, in 5D EMCS theory, an asymptotically flat stationary rotating charged non-extremal black hole solution with multiple disconnected horizons that is regular on and outside all the event horizons. If (1) the black hole space-time admits, besides the stationary Killing vector, two mutually commuting axial Killing vector fields and (2) the topology of each horizon is either $S^{3}$ or $S^{1}\times S^{2}$, then the black hole space-time is uniquely characterized by its rod structure, asymptotic charges as well as the local charges and fluxes.} \\ \\
In order to prove the above theorem we will take the same approach
as the one taken in \cite{Tomizawa:2009ua}, consisting in showing
that the LHS of the Mazur Identity vanishes on the boundary
$\partial \Sigma$. We start by rewriting the LHS of \eqref{E:Mi} as:

\begin{equation} \label{E:Mis}
\int_{\partial \Sigma} r \partial_{a} tr{\Psi} dS^{a}=\sum_{I }\int_{I} r \partial_{z} tr{\Psi} dz + \int_{\partial \Sigma_{\infty}} r \partial_{a} tr{\Psi} dS^{a}
\end{equation}
where the index $I$ denotes a specific rod.\\ \\
In order to show that the RHS of the equation above vanishes we will need to derive the boundary conditions for a field configuration
$\Theta=\{\lambda_{ab}, \omega_{a}, \mu, \psi_{a}\}$. The quantities which need to be specified to appropriately derive these boundary conditions for each rod $I$ and at infinity depend on the result of the calculation of the RHS of equation \eqref{E:Mis}. Computing explicitly the RHS of equation \eqref{E:Mis} one can prove the following lemma:\\ \\
\textbf{Lemma 1:}  \emph{In the context of Theorem 1, two solutions of the equations of motion $\Theta_{0}$ and $\Theta_{1}$ with the same rod structure and the same asymptotic charges are isometric if and only if at each rod $I = [\kappa_{i},\kappa_{i+1}]$ the value of the potentials $\{\omega_{a}, \mu, \psi_{a}\}$ are the same for the two solutions at both of the rod end points $\kappa_{i},\kappa_{i+1}$.} \\ \\
This means that to characterize a solution one needs to define the local charges and fluxes such that they fully determine the potentials $\{\omega_{a}, \mu, \psi_{a}\}$ at all of the rod end points.\\ \\
We will now define the rod structure of the class of solutions we are
interested in and show that the values of the potentials $\{\omega_{a}, \mu, \psi_{a}\}$ are determined from the rod structure, asymptotic charges and the local charges and fluxes.

\subsection{Rod Structure of the General Black Hole Solution} \label{grod}

The most general black hole solution in this theory obeying our two requirements above corresponds to a charged 2-spin Myers-Perry black hole with $n_{1}$ 2-spin concentric charged dipole black rings placed on the $\psi$ orthogonal plane and $n_{2}$ 2-spin concentric charged dipole black rings placed on the $\phi$ orthogonal plane. \\ \\ This has the following rod structure:

\begin{figure}[H]
\centering
\includegraphics[width=1.06\textwidth, height=0.07\textheight]{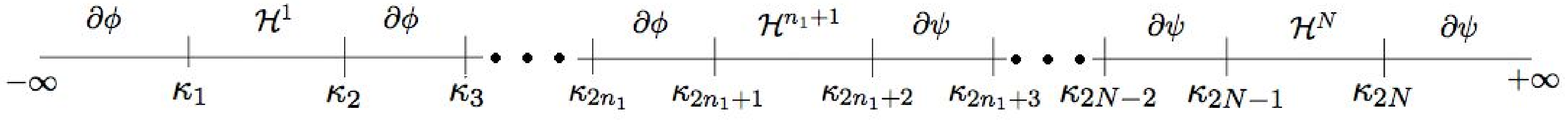}
\end{figure}

Here $n_{1}$ is the total number of concentric black rings with the $S^{1}$ parameterized by $\psi$. $N$ is the total number of black hole horizons and is defined as $N=n_{1}+n_{2}+1$, where $n_{2}$ is the total number of concentric black rings with the $S^{1}$ parameterized by $\phi$ and the extra factor of $1$ accounts for the Myers-Perry black hole.\\ \\
Since we can rescale and shift the $z$-axis without changing the properties of the solution we can always define dimensionless rod structure parameters in the form above and satisfying:
\[
0\leq\kappa_{1}<\kappa_{2}<...<\kappa_{2N}\leq1
\]
Define $i$ as an index that runs over $i=1,...,N$, where $i=m$ labels the Myers-Perry black hole, then we can summarize the rod structure above as \footnote{For convenience we add an extra boundary to the rod structure: the boundary at infinity.}:\\ \\
\emph{(i)}  $\phi$-invariant plane: $\Sigma_{\phi}=\{(r,z)|r=0,-\infty<z<\kappa_{1}\}$ and rod vector $v=(0,1,0)$.\\ \\
\emph{(ii)} $\phi$-invariant plane: $n_{1}$ rods with $\Sigma_{\phi^{i}}=\{(r,z)|r=0, \kappa_{i}<z<\kappa_{i+1},~i=2l,~1\leq l\leq n_{1}\}$ and rod vector $v=(0,1,0)$. \\ \\
 \emph{(iii)} BR Horizon: $n_{1}$ rods with $\Sigma_{\mathcal{H}^{i}}=\{(r,z)|r=0, \kappa_{i}<z<\kappa_{i+1},~i=2l-1,~1\leq l\leq n_{1}\}$ and rod vector $v=(1,\Omega^{i}_{\phi},\Omega^{i}_{\psi})$ where the  $S^{1}$ is parameterized by $\psi$.\\ \\
\emph{(iv)} BH Horizon: $\Sigma_{\mathcal{H}^{m}}=\{(r,z)|r=0, \kappa_{2n_{1}+1}<z<\kappa_{2n_{1}+2}\}$ and rod vector $v=(1,\Omega^{m}_{\phi},\Omega^{m}_{\psi})$.\\ \\
 \emph{(v)} BR Horizon: $n_{2}$ rods with $\Sigma_{\mathcal{H}^{i}}=\{(r,z)|r=0, \kappa_{i}<z<\kappa_{i+1},~i=2l-1,~(n_{1}+2)\leq l\leq N\}$ and rod vector $v=(1,\Omega^{i}_{\phi},\Omega^{i}_{\psi})$  where the $S^{1}$ is parameterized by $\phi$.\\ \\
  \emph{(vi)} $\psi$-invariant plane: $n_{2}$ rods with $\Sigma_{\psi^{i}}=\{(r,z)|r=0, \kappa_{i}<z<\kappa_{i+1},~i=2l,~(n_{1}+2) \leq l\leq (N-1)\}$ and rod vector $v=(0,0,1)$. \\ \\
 \emph{(vii)} $\psi$-invariant plane: $\Sigma_{\psi}=\{(r,z)|r=0,\kappa_{2N}<z<+\infty\}$ with rod vector $v=(0,0,1)$. \\ \\
 \emph{(viii)} Asymptotic infinity: $\Sigma_{\infty}=\{(r,z)|\sqrt{r^{2}+z^{2}}\to\infty, \frac{z}{\sqrt{r^{2}+z^{2}}}=const\}$\\

 \subsection{Boundary Value Problem}

Here we provide the necessary tools that will allow us to set boundary conditions on the different fields $\{\lambda_{ab}, \omega_{a}, \mu, \psi_{a}\}$ at the rods and at infinity. We also define intrinsic local charges and fluxes that will be determinant in the proof of our uniqueness theorem. These local charges will measure the intrinsic angular momenta, electric charge, Maxwell electric charge, dipole charge and Chern-Simons dipole charge of a specific horizon $\mathcal{H}^{i}$, where the fluxes will measure the magnetic flux and the Chern-Simons flux at each fixed plane of rotation $\phi^{k},\psi^{k}$.\\ \\
\textbf{I. Metric Fields $\lambda_{ab}$ }\\ \\
To impose boundary conditions on these fields one needs to write down the metric near a rod. It is then useful to make use of the following theorem \cite{Harmark:2004rm}:\\ \\
\textbf{Theorem 2:} \emph{Consider a rod $[z_{1},z_{2}]$ for a solution $G(r,z)$. Then we can find an orthogonal matrix $\Lambda_{*}$ such that the solution $\tilde{G}(r,z)=\Lambda_{*}^{T}G(r,z)\Lambda_{*}$ has the property that $\tilde{G}_{1i}(r,z)=0$ for $i=1,...D-2$ and $z ~\epsilon ~[z_{1},z_{2}]$.}\\ \\
This implies that we can write the metric near a rod as:

\begin{equation} \label{E:dsrod}
ds^{2}=\Sigma_{ij}A_{ij}(z)dx^{i}dx^{j} + a(z)[r^{2}(dx^{1})^{2} + c^{2}(dr^{2}+dz^{2})], ~r\to0
\end{equation}
where $A_{ij}(z)$ and $a(z)$ are functions that depend on the particular solution and c is a positive constant. With this the fields $\lambda_{ab}$ are specified near a rod. At infinity it is straightforward to use the asymptotic expansion of the metric in appendix \ref{ac} and the map on appendix \ref{aa}.\\ \\ \\
\textbf{II. Electric Potentials $\psi_{a}$ }\\ \\
The gauge field $A$, and hence the electric potentials, can be specified by writing down the field strength $F$ near a rod. This can easily be done through the equation of motion \eqref{E:mot1}. In fact, considering for example the rod corresponding to the $\phi$-invariant plane, one can find the following relations:
\begin{equation} \label{E:f1}
F_{\phi r}^{2}+F_{\phi z}^{2}=r^{2}H_{\phi}(z)
\end{equation}
\begin{equation} \label{E:f2}
F_{\psi r}^{2}+F_{\psi z}^{2}=H_{\psi}(z)
\end{equation}
\begin{equation} \label{E:f3}
F_{t r}^{2}+F_{t z}^{2}=H_{t}(z)
\end{equation}
where $H_{\phi}(z),~H_{\psi}(z),~H_{t}(z)$ are non-trivial functions of $z$. Nevertheless, this is not enough to specify completely the potentials $\psi_{a}$ in the presence of a dipole charge. This charge is defined as:
\begin{equation} \label{E:dc}
q_{a}=\frac{1}{2\pi}\int_{S^{2}}F
\end{equation}
where the integral is performed over an $S^{2}$ that encloses the ring once and the $S^{1}$ is parameterized by $a=(\psi,\phi)$. Besides having to specify an $S^{2}$ it is also necessary to specify a tangent vector along the ring, hence diametrically opposite points of the ring will have opposite charges \cite{Emparan:2001wk}.\\ \\
Bearing in mind the rod structure presented in section \ref{grod} and using the definition $F=dA$ we obtain for a ring with an $S^{1}$ parameterized by $\psi$:
\begin{equation} \label{E:dcc1}
q_{\psi}=\frac{1}{2\pi}\int_{0}^{2\pi}\int_{z_{1}}^{z_{2}}\partial_{z} A_{\phi}dz d\phi=[A_{\phi}(0,z_{2}) - A_{\phi}(0,z_{1})] = [A_{\phi}]_{I}
\end{equation}
where $z_{1},z_{2}$ are the endpoints of a specific rod interval $I$ and $A$ is a function of $(r,z)$ only.
For a ring placed on the
other orthogonal plane we find \footnote{There is a small subtlety
that one has to consider while parameterizing the 2-sphere for a ring
placed on the $\phi$ orthogonal plane. This subtlety is manifest
when we define the coordinates $cos\theta=-1+\frac{\kappa_{2N}}{z}$
at the rightmost rod representing the $\psi$-invariant plane. Hence,
one needs to take into account the minus sign coming from the change
in orientation. }\label{foot}:
\begin{equation} \label{E:dcc2}
q_{\phi}=-[A_{\psi}]_{I}
\end{equation}
We can then easily impose boundary conditions on these potentials by defining the intrinsic dipole charge measured at each horizon:
\begin{equation} \label{E:id}
q^{i}_{a}=\frac{1}{2\pi}\int_{\mathcal{H}^{i}}F
\end{equation}
Then, imagine starting from the leftmost rod representing the
$\phi$-fixed plane of rotation. Here $A_{\phi}=0$ since from
equation \eqref{E:epot} $\psi_{\phi}$ is constant on any
$\phi$-fixed plane and, as we will see below, $A_{\phi}\to0$ at
infinity. Hence, until we hit the Myers-Perry black hole we can
specify the value of $\psi_{\phi}$ at any $\phi^{k}$-fixed plane of
rotation by adding the contributions from each black ring horizon:
\begin{equation} \label{E:vg}
\hat{q}^{k}_{\psi}=\psi_{\phi}^{k}=\frac{1}{\sqrt{3}}\sum_{i=1}^{k}[A_{\phi}]_{\mathcal{H}^{i}}=\frac{1}{\sqrt{3}}\sum_{i=1}^{k} q^{i}_{\psi}
\end{equation}
and, starting from the rightmost rod, for any $\psi^{k}$-fixed plane
of
rotation we obtain a similar expression for  $\psi_{\psi}^{k}$. \\ \\
The intrinsic dipole charges and the considerations on the gauge field above are sufficient to set correct boundary conditions on the potentials $\psi_{a}$. This potential can be specified at infinity through the equation of motion \eqref{E:mot1} as above. \\ \\ \\
\textbf{III. Twist Potentials $\omega_{a}$ }\\ \\
To impose restrictions on these potentials we start with equation \eqref{E:twi} and apply Stoke's theorem in order to obtain a relation between the value of the twist potentials on the rods and on the boundary at infinity:

\begin{equation} \label{E:twist}
\int_{\kappa_{1}}^{\kappa_{2N}}\omega_{a,z}dz= \int_{\partial\Sigma_{\infty}}W_{a} + \int_{\partial\Sigma_{\infty}}\psi_{a}(3d\mu + \epsilon^{bc}\psi_{b}d\psi_{c})
\end{equation}
Now, since the second term on the RHS of \eqref{E:twist} vanishes, as we will see when considering the asymptotic behavior of the solution, and the twist potentials are invariant under the action of the 2-independent rotation isometries then the first term on the RHS above is seen to be proportional to the total angular momenta measured at infinity, i.e.:
\begin{equation} \label{E:twist2}
\int_{\kappa_{1}}^{\kappa_{2N}}\omega_{a,z}dz=\alpha J^{T}_{a}
\end{equation}
with $J^{T}_{a}$ given by:
\begin{equation} \label{E:tj}
J^{T}_{a}=\frac{1}{16\pi}\int_{S^{3}_{\infty}}*dV_{(a)}
\end{equation}
Due to the fact that we have the freedom to add constants to the
twist potentials we will choose for convenience the value
$\alpha=\frac{4}{\pi}$. Hence we can set the value
$\omega_{a}=-\frac{2J^{T}_{a}}{\pi}$ and
$\omega_{a}=\frac{2J^{T}_{a}}{\pi}$ on the leftmost and rightmost
rods respectively since the twist potentials are constant there. It
remains to inspect the LHS of equation \eqref{E:twist}. A closer
look at equations \eqref{E:epot}-\eqref{E:v} tells us that in the
presence of a dipole charge the twist potentials do not necessarily
vanish on the fixed planes of rotation. It is thus convenient to
split this integral into three different parts:
\begin{equation} \label{E:twist3}
\int_{\kappa_{1}}^{\kappa_{2N}}\omega_{a,z}dz=\sum^{N}_{i=1}J^{i}_{a} -  2\sum^{n_{1}}_{i=1}\int_{\partial\Sigma_{\phi^{i}}}\psi_{a}\psi_{\phi}d\psi_{\psi} + 2\sum^{N-1}_{i=n_{1}+1}\int_{\partial\Sigma_{\psi^{i}}}\psi_{a}\psi_{\psi}d\psi_{\phi}
\end{equation}
where we have defined the intrinsic angular momenta measured near each horizon by:
\begin{equation} \label{E:inj}
J^{i}_{a}=\int_{\partial\Sigma_{\mathcal{H}^{i}}}W_{a} + \int_{\partial\Sigma_{\mathcal{H}^{i}}} \psi_{a}(3d\mu + \epsilon^{bc}\psi_{b}d\psi_{c})
\end{equation}
In the above expression the first term on the RHS is proportional to
the angular momenta Komar integral evaluated on the horizon while
the second term accounts for the electromagnetic contribution to the
momenta. We now focus on the second term of equation
\eqref{E:twist3}, which is in essence the same as the third term. On
a specific $\phi^{k}$-fixed plane of rotation we have for $a=\phi$
the quantity:
\begin{equation} \label{E:mff}
\int_{\partial\Sigma_{\phi^{k}}}(\psi_{\phi}^{k})^{2}d\psi_{\psi}=(\hat{q}_{\psi}^{k})^{2}\int_{z_{1}}^{z_{2}}\partial_{z}\psi_{\psi}dz=(\hat{q}_{\psi}^{k})^{2}[\psi_{\psi}]_{I}=\frac{1}{2\pi\sqrt{3}}(\hat{q}_{\psi}^{k})^{2}\Phi^{k}_{\phi}
\end{equation}
In the second step we have used the definition \eqref{E:vg} and the
fact that $\psi_{\phi}$ is constant on any $\phi$-fixed plane of
rotation. In the last step we have used the definition of the
magnetic flux measured on each $\phi^{k}$-fixed plane of rotation \footnote{As in the case of the dipole charge, one needs to take into account the minus sign coming from the change in orientation on the $\psi^{k}$-fixed planes of rotation in both the magnetic flux and the Chern-Simons flux which will be defined below.}\label{foot1}:
\begin{equation} \label{E:mf}
\Phi^{k}_{\phi}=\int_{C^{k}}F
\end{equation}
where the 2-dimensional surface $C^{k}$ has the topology of a disk
with a hole in the middle. If we now take $a=\psi$ we obtain:
\begin{equation} \label{E:csff}
\int_{\partial\Sigma_{\phi^{k}}}\psi_{\phi}^{k}\psi_{\psi}d\psi_{\psi}=\hat{q}_{\psi}^{k}\int_{z_{1}}^{z_{2}}\psi_{\psi}\partial_{z}\psi_{\psi}dz=\frac{1}{2}\hat{q}_{\psi}^{k}[(\psi_{\psi})^{2}]_{I}=-\frac{1}{6\pi}\hat{q}_{\psi}^{k}\Xi_{\phi}^{k}
\end{equation}
where we have defined the Chern-Simons flux measured on
each $\phi$-fixed plane of rotation by:%
\footnote{We write here Chern-Simons flux since $\Xi^k_\phi$ clearly resembles a flux. However, we have not examined the physical properties of $\Xi^k_\phi$ in detail. This would be interesting to study further. The same goes for the Chern-Simons dipole charge that we will introduce below.}\label{foot3}
\begin{equation} \label{E:csf}
\Xi_{\phi}^{k}=\int_{C^{k}}A\wedge (i_{V_{(2)}}F)
\end{equation}
Now, following the analysis of \cite{Tomizawa:2009tb} we can write for any  $\phi^{k}$-fixed plane:
\begin{equation} \label{E:gfphi}
\psi_{\psi}^{k}=f^{k}(z)+O(r^{2})
\end{equation}
Then, using equations \eqref{E:epot}-\eqref{E:v} we can determine the potentials $\omega^{k}_{a}$ as functions of $\psi_{\psi}^{k}$,
\begin{equation} \label{E:gfphi1}
\omega_{\phi}^{k}=-2(\hat{q}_{\psi}^{k})^{2} \psi_{\psi}^{k} + c_{\phi}^{k},~\omega_{\psi}^{k}=-\hat{q}_{\psi}^{k} (\psi_{\psi}^{k})^{2} + c_{\psi}^{k}
\end{equation}
where $c_{\phi}^{k},~c_{\psi}^{k}$ are constants to be determined.
To determine these constants, as it has been shown in \cite{Tomizawa:2009tb} for the case of a single dipole ring, it is necessary to have the knowledge of the function $f^{k}(z)$ at one of the endpoints of the rod interval we are considering as well as the knowledge of the potentials $\omega^{k}_{a}$ at that same point. The knowledge of the last we can easily obtain, it is the sum of the contributions of the intrinsic angular momenta and fluxes given by:
\begin{equation} \label{E:twistk}
\omega_{\psi}^{k}(\kappa_{i})=-\frac{2J^{T}_{\psi}}{\pi} + \sum_{i=1}^{k}J_{\psi}^{i} + \frac{1}{3\pi}\sum_{i=1}^{k-1}\hat{q}_{\psi}^{i}\Xi_{\phi}^{i}
\end{equation}
while for the latter we can make use of the definitions of magnetic flux and Chern-Simons flux \eqref{E:mf}, \eqref{E:csf} to obtain the following two equations for any $\phi^{k}$-fixed plane with $z~\epsilon~[\kappa_{i},\kappa_{i+1}]$:
\begin{equation} \label{E:const1}
f^{k}(\kappa_{i+1}) - f^{k}(\kappa_{i}) = \frac{\Phi^{k}_{\phi}}{2\pi\sqrt{3}}
\end{equation}
\begin{equation} \label{E:const2}
f^{k}(\kappa_{i+1})^{2} - f^{k}(\kappa_{i})^{2} =- \frac{\Xi^{k}_{\phi}}{3\pi}
\end{equation}
For which we obtain a unique solution:
\begin{equation} \label{E:const3}
f^{k}(\kappa_{i})=-\frac{1}{\sqrt{3}}(\frac{\Xi^{k}_{\phi}}{\Phi^{k}_{\phi}} + \frac{\Phi^{k}_{\phi}}{4\pi})
\end{equation}
Then, the constants $c_{\phi}^{k},~c_{\psi}^{k}$ are easily obtained. As an example one can write $c^{k}_{\psi}$ as:
\begin{equation} \label{E:constpsi}
c^{k}_{\psi}=\omega^{k}_{\psi}(\kappa_{i})  + \frac{1}{3} \hat{q}^{k}_{\psi} (\frac{\Xi^{k}_{\phi}}{\Phi^{k}_{\phi}} + \frac{\Phi^{k}_{\phi}}{4\pi})^{2}
\end{equation}
The same analysis can be done for any $\psi^{k}$-fixed plane.\\ \\
The expression \eqref{E:constpsi} is not valid for the case in which the flux $\Phi^{k}_{\phi}$ vanish. In fact, for this case the fluxes are not enough to determine the function $f^{k}(\kappa_{i})$ since from equations \eqref{E:const1} and \eqref{E:const2} we only obtain $f^{k}(\kappa_{i})=f^{k}(\kappa_{i+1})$. At the end of the section below we will explain how to deal with this case.\\ \\
The considerations above are enough to specify the boundary conditions for these potentials on any rod for which $\Phi^{k}_{\phi}\ne 0$. On the boundary at infinity we can use the asymptotic metric expansion and the equations \eqref{E:atr}-\eqref{E:v}. \\ \\ \\
\textbf{IV. Magnetic Potential $\mu$ }\\ \\
In a similar fashion as above we start with equation \eqref{E:u} and apply Stoke's Theorem in order to relate the difference in the magnetic potential between the leftmost and rightmost horizon rod with the total electric charge in a very simple way:
\begin{equation} \label{E:u2}
\int_{\kappa_{1}}^{\kappa_{2N}}\mu_{,z}dz=\int_{\kappa_{1}}^{\kappa_{2N}}\frac{1}{\sqrt{3}}[\frac{\tau}{r}(A_{t,r}-a^{\phi}_{t}A_{\phi,r}-a^{\psi}_{t}A_{\psi,r}) - \frac{1}{\sqrt{3}}(A_{\phi}A_{\psi,z}-A_{\psi}A_{\phi,z})]dz=\frac{4}{\sqrt{3}\pi} Q^{T}
\end{equation}
where the total electric charge $Q^{T}$ is given by:
\begin{equation} \label{E:qcharge}
Q^{T}=\frac{1}{16\pi}\int_{S^{3}_{\infty}}(*F + \frac{1}{\sqrt{3}}A \wedge F)
\end{equation}
Hence we can set the value $\mu=-\frac{2Q^{T}}{\sqrt{3}\pi}$ and
$\mu=\frac{2Q^{T}}{\sqrt{3}\pi}$ on the leftmost and rightmost rods
respectively. Again, in the presence of dipole charges, the magnetic
potential is not necessarily constant on the fixed planes of
rotation. Hence we can write the LHS of equation \eqref{E:u2} into a
sum of integrals over the horizon and axes rods:
\begin{equation} \label{E:u22}
\int_{\kappa_{1}}^{\kappa_{2N}}\mu_{,z}dz=\frac{4}{\sqrt{3}\pi} \sum_{i=1}^{N}Q^{i} - \sum_{i=1}^{n_{1}}\int_{\partial\Sigma_{\phi^{i}}}\psi_{\phi}d\psi_{\psi} + \sum_{i=n_{1}+1}^{N-1}\int_{\partial\Sigma_{\psi^{i}}}\psi_{\psi}d\psi_{\phi}
\end{equation}
where we have defined the intrinsic electric charge measured near each horizon by:
\begin{equation} \label{E:ic}
Q^{i}=\frac{1}{16\pi}\int_{\mathcal{H}^{i}}(*F + \frac{1}{\sqrt{3}}A \wedge F)
\end{equation}
The second term on the RHS of \eqref{E:u22} can be rewritten on an
fixed plane of rotation $\phi^{k}$ as above:
\begin{equation} \label{E:uf}
\int_{\partial\Sigma_{\phi^{k}}}\psi_{\phi}^{k}d\psi_{\psi}=\hat{q}_{\psi}^{k}\int_{z_{1}}^{z_{2}}\partial_{z}\psi_{\psi}dz=\frac{1}{2\pi\sqrt{3}}\hat{q}_{\psi}^{k}\Phi^{k}
\end{equation}
 proceed as above and use equation \eqref{E:u} to write:
\begin{equation} \label{E:muf}
\mu^{k}=-\hat{q}^{k}_{\psi}\psi_{\psi}^{k} + c_{\mu}^{k}
\end{equation}
where $c_{\mu}^{k}$ is a constant which can be determined as in the previous case for $\Phi^{k}_{\phi}\ne 0$ by specifying the value of $\mu^{k}$ at one of the rod endpoints:
\begin{equation} \label{E:uk}
\mu^{k}(\kappa_{i})=-\frac{2Q^{T}}{\sqrt{3}\pi} + \frac{4}{\sqrt{3}\pi} \sum_{i=1}^{k}Q^{i}  - \frac{1}{2\pi\sqrt{3}}\sum_{i=1}^{k-1}\hat{q}_{\psi}^{i}\Phi^{i}
\end{equation} \\
We will now consider the case $\Phi^{k}_{\phi}=0$ in detail. In this case the problem relies on finding for each rod $I$ the value of $f^{k}(\kappa_{i})$. This can be done as follows. Suppose that at any $\phi^{k}$-fixed plane $\Phi^{k}_{\phi}=0$, then consider the ring horizon rod $\mathcal{H}^{i}$ at the left of the $\phi^{k}$-fixed plane. For this rod we have defined the electric charge $Q^{i}$ given by \eqref{E:ic}. However, in theories with Chern-Simons terms there are different notion of charges and the electric charge $Q^{i}$ is known as the Page charge \cite{Marolf:2000cb}. If we now in addition define the Maxwell charge for this horizon as:
\begin{equation} \label{E:im}
Q^{i}_{M}=\frac{1}{16\pi}\int_{\mathcal{H}^{i}}*F
\end{equation}
Then, the Chern-Simon contribution to the electric charge is given by:
\begin{equation} \label{E:ics}
Q^{i}-Q^{i}_{M}=\frac{1}{16\pi \sqrt{3}}\int_{\mathcal{H}^{i}}A \wedge F
\end{equation}
If we now take the RHS of the equation above and express it in terms of the twist potentials we obtain:
\begin{equation} \label{E:ics1}
\int_{\mathcal{H}^{i}}A \wedge F=\frac{3\pi}{4}\int_{\kappa_{i}}^{\kappa_{i+1}}(\psi_{\psi}\partial_{z}\psi_{\phi}-\psi_{\phi}\partial_{z}\psi_{\psi})dz=\frac{3\pi}{4}([\psi_{\psi}\psi_{\phi}]_{I} - 2\int_{\kappa_{i}}^{\kappa_{i+1}}\psi_{\phi}\partial_{z}\psi_{\psi}dz)
\end{equation}
The second term on the RHS side we can obtain by defining the Chern-Simons dipole charge:
\begin{equation} \label{E:dcs}
\mathcal{Q}^{i}_{\psi}=\frac{1}{2\pi}\int_{\mathcal{H}^{i}}A\wedge(i_{V_{(2)}}F)
\end{equation}
As with the dipole charge we have to keep $\psi$ constant and specify a tangent vector along the ring. Using this definition and equation \eqref{E:ics1} we obtain, setting $i=k$, a recursive relation for $f^{k}(z)$ at one of the endpoints:
\begin{equation} \label{E:fko}
\hat{q}_{\psi}^{k}f^{k}(\kappa_{k+1})=\frac{4}{\sqrt{3}\pi}(Q^{k}-Q^{k}_{M}) - \frac{2}{3}\mathcal{Q}^{k}_{\psi} + \hat{q}_{\psi}^{k-1}f^{k-1}(\kappa_{i})
\end{equation}
This can always be exactly determined since $f^{k-1}(\kappa_{i})$ can be determined either by the fluxes or using the above relation for the previous horizon, noting that at the leftmost semi-infinite rod $\hat{q}_{\psi}^{k-1}=0$. This can then be used to obtain the constants $c^{k}_{\phi},~c^{k}_{\psi},~c^{k}_{\mu}$ as above. The same analysis can be carried out at any $\psi^{k}$-fixed plane. However, note that expression \eqref{E:fko} cannot be used to determine $f^{k}(\kappa_{k+1})$ when $\hat{q}_{\psi}^{k}=0$, nevertheless this is not necessary in this case since by looking at equations \eqref{E:gfphi1} and \eqref{E:muf} the constants $c^{k}_{\phi},~c^{k}_{\psi},~c^{k}_{\mu}$ are immediately determined.\\ \\
Summarizing, in addition to the local angular momenta $J_{a}^{i}$ and the electric charges $Q^{i}, Q^{i}_{M}$, we need in general to specify the charges:
\[
q^{i}_{\psi}=\frac{1}{2\pi}\int_{\mathcal{H}^{i}}F,~\mathcal{Q}^{i}_{\psi}=\frac{1}{2\pi}\int_{\mathcal{H}^{i}}A\wedge(i_{V_{(2)}}F)
\]
at the horizon of each ring and the fluxes:
\[
\Phi^{k}_{\phi}=\int_{C^k}F,~\Xi^{k}_{\phi}=\int_{C^k}A\wedge(i_{V_{(2)}}F)
\]
at each $\phi^k$-fixed plane of rotation, and similarly for all the rings placed at the other orthogonal plane and for each $\psi^k$-fixed plane.\\ \\
We note that all known regular analytical solutions with multiple disconnected horizons fall into the class with $\Phi^{k}_{a}=0$, $\mathcal{Q}^{i}_{a}=0$ and $Q^{i}-Q^{i}_{M}=0$.\\ \\
These intrinsic charges and fluxes are sufficient to specify all boundary conditions on the potentials $\mu$. At infinity we use the same approach as we did for the twist potentials. \\

\subsection{Proof of the Uniqueness Theorem} \label{proof}

In this section we will apply the above considerations to the rod structure of section \ref{grod}. In order to prove our uniqueness theorem we will need to compute the quantity $r\partial_{z}Tr\Psi$ on the LHS of the Mazur identity \eqref{E:Mi} and show that it vanishes as $r\to 0$ on all rods and at infinity.\\ \\
\emph{(i)}  $\phi$-invariant plane: $\Sigma_{\phi}=\{(r,z)|r=0,-\infty<z<\kappa_{1}\}$ and rod vector $v=(0,0,1)$.\\
\[
\lambda_{\phi \phi}=\mathcal{O}(r^{2}), ~\lambda_{\psi \psi}=\mathcal{O}(1), ~\lambda_{\phi \psi}=\mathcal{O}(r^{2})
\]
\[
\psi_{\phi}=\mathcal{O}(r^{2}), ~\psi_{\psi}=\mathcal{O}(1)
\]
\[
\omega_{a}=-\frac{2J^{T}_{a}}{\pi}+\mathcal{O}(r^{2}),~\mu=-\frac{2Q^{T}}{\sqrt{3\pi}} + \mathcal{O}(r^{2})
\]
\[
r\partial_{z}tr(\Psi)=\mathcal{O}(r)
\]
\emph{(ii)} $\phi$-invariant plane: $n_{1}$ rods with $\Sigma_{\phi^{k}}=\{(r,z)|r=0, \kappa_{i}<z<\kappa_{i+1},~i=2l,~1\leq l\leq n_{1}\}$ and rod vector $v=(0,0,1)$. \\
\[
\lambda_{\phi \phi}=\mathcal{O}(r^{2}), ~\lambda_{\psi \psi}=\mathcal{O}(1), ~\lambda_{\phi \psi}=\mathcal{O}(r^{2})
\]
\[
\psi_{\phi}^{k}=\hat{q}_{\psi}^{k} + \mathcal{O}(r^{2}), ~\psi_{\psi}^{k}=f^{k}(z) + \mathcal{O}(r^{2}) 
\]
\[
\omega_{\phi}^{k}=-2(\hat{q}_{\psi}^{k})^{2} f^{k}(z) + c_{\phi}^{k} + \mathcal{O}(r^{2}),~\omega_{\psi}^{k}=-\hat{q}_{\psi}^{k} (f^{k}(z))^{2} + c_{\psi}^{k} + \mathcal{O}(r^{2})
\]
\[
\mu^{k}=-\hat{q}^{k}_{\psi}f^{k}(z) + c_{\mu}^{k} + \mathcal{O}(r^{2})
\]
\[
r\partial_{z}tr(\Psi)=\mathcal{O}(r)
\]
if $\Phi^{k}_{\phi}\ne0\vee(\Phi^{k}_{\phi}=0\wedge\hat{q}_{\psi}^{k}=0)$,
\[
c_{\phi}^{k}=-\frac{2J^{T}_{\phi}}{\pi} + \sum_{i=1}^{k}J_{\phi}^{i} - \frac{1}{\pi\sqrt{3}}\sum_{i=1}^{k}(\hat{q}_{\psi}^{i})^{2}\Phi^{i}_\phi  -\frac{2}{\sqrt{3}}(\hat{q}^{k}_{\psi})^{2} \left(\frac{\Xi^{k}_{\phi}}{\Phi_\phi^{k}} + \frac{\Phi_\phi^{k}}{4\pi} \right)
\]
\[
c_{\psi}^{k}=-\frac{2J^{T}_{\psi}}{\pi} + \sum_{i=1}^{k}J_{\psi}^{i} + \frac{1}{3\pi}\sum_{i=1}^{k-1}\hat{q}_{\psi}^{i}\Xi_{\phi}^{i} + \frac{1}{3}\hat{q}^{k}_{\psi} \left(\frac{\Xi^{k}_{\phi}}{\Phi_\phi^{k}} + \frac{\Phi_\phi^{k}}{4\pi} \right)^{2}
\]
\[
c_{\mu}^{k}=-\frac{2Q^{T}}{\sqrt{3}\pi} + \frac{4}{\sqrt{3}\pi} \sum_{i=1}^{k}Q^{i}  - \frac{1}{2\pi\sqrt{3}}\sum_{i=1}^{k-1}\hat{q}_{\psi}^{i}\Phi_\phi^{i} - \frac{1}{\sqrt{3}} \hat{q}^{k}_{\psi} \left(\frac{\Xi^{k}_{\phi}}{\Phi_\phi^{k}} + \frac{\Phi_\phi^{k}}{4\pi}\right)
\]
if $\Phi^{k}_{\phi}=0~\wedge~\hat{q}_{\psi}^{k}\ne0$,
\[
c_{\phi}^{k}=-\frac{2J^{T}_{\phi}}{\pi} + \sum_{i=1}^{k}J_{\phi}^{i} - \frac{1}{\pi\sqrt{3}}\sum_{i=1}^{k}(\hat{q}_{\psi}^{i})^{2}\Phi_\phi^{i}  + 2\hat{q}^{k}_{\psi} \left(\frac{4}{\sqrt{3}\pi}(Q^{k}-Q^{k}_{M})-\frac{2}{3}\mathcal{Q}^{k}_{\psi}+\hat{q}^{k}_{\psi}f^{k-1}(\kappa_{i})\right)
\]
\[
c_{\psi}^{k}=-\frac{2J^{T}_{\psi}}{\pi} + \sum_{i=1}^{k}J_{\psi}^{i} + \frac{1}{3\pi}\sum_{i=1}^{k-1}\hat{q}_{\psi}^{i}\Xi_{\phi}^{i} + \frac{1}{\hat{q}^{k}_{\psi}}\left(\frac{4}{\sqrt{3}\pi}(Q^{k}-Q^{k}_{M})-\frac{2}{3}\mathcal{Q}^{k}_{\psi}+\hat{q}^{k}_{\psi}f^{k-1}(\kappa_{i})\right)^2
\]
\[
c_{\mu}^{k}=-\frac{2Q^{T}}{\sqrt{3}\pi} + \frac{4}{\sqrt{3}\pi} \sum_{i=1}^{k}Q^{i}  - \frac{1}{2\pi\sqrt{3}}\sum_{i=1}^{k-1}\hat{q}_{\psi}^{i}\Phi_\phi^{i} +\frac{4}{\sqrt{3}\pi}(Q^{k}-Q^{k}_{M})-\frac{2}{3}\mathcal{Q}^{k}_{\psi}+\hat{q}^{k}_{\psi}f^{k-1}(\kappa_{i})
\]
\emph{(iii), (iv), (v)} horizon rods:\\
\[
\lambda_{ab}=\mathcal{O}(1), ~ \omega_{a}=\mathcal{O}(1)
\]
\[
\mu=\mathcal{O}(1),~\psi_{a}=\mathcal{O}(1)
\]
\[
r\partial_{z}tr(\Psi)=\mathcal{O}(r)
\]
\emph{(vi)} $\psi$-invariant plane: $n_{2}$ rods with $\Sigma_{\psi^{i}}=\{(r,z)|r=0, \kappa_{i}<z<\kappa_{i+1},~i=2l,~(n_{1}+2) \leq l\leq (N-1)\}$ and rod vector $v=(0,0,1)$. \\
\[
\lambda_{\phi \phi}=\mathcal{O}(1), ~\lambda_{\psi \psi}=\mathcal{O}(r^{2}), ~\lambda_{\phi \psi}=\mathcal{O}(r^{2})
\]
\[
\psi_{\phi}^{k}=f^{k}(z) + \mathcal{O}(r^{2}), ~\psi_{\psi}^{k}=\hat{q}_{\phi}^{k} + \mathcal{O}(r^{2})
\]
\[
\omega_{\phi}^{k}=2\hat{q}_{\phi}^{k}(f^{k}(z))^2 + c_{\phi}^{k} + \mathcal{O}(r^{2}),~\omega_{\psi}^{k}=(\hat{q}_{\phi}^{k})^{2} f^{k}(z) + c_{\psi}^{k} + \mathcal{O}(r^{2})
\]
\[
\mu^{k}=\hat{q}^{k}_{\phi}f^{k}(z) + c_{\mu}^{k} + \mathcal{O}(r^{2})
\]
\[
r\partial_{z}tr(\Psi)=\mathcal{O}(r)
\]
if $\Phi^{k}_{\psi}\ne0\vee(\Phi^{k}_{\psi}=0\wedge\hat{q}_{\psi}^{k}=0)$,
\[
c_{\phi}^{k}=\frac{2J^{T}_{\phi}}{\pi} - \sum_{i=k+1}^{N}J_{\phi}^{i} - \frac{1}{3\pi}\sum_{i=k}^{N-1}\hat{q}_{\phi}^{i}\Xi_{\psi}^{i} - \frac{2}{3}\hat{q}^{k}_{\phi} \left(\frac{\Xi^{k}_{\psi}}{\Phi_\psi^{k}} + \frac{\Phi_\psi^{k}}{4\pi} \right)^{2}
\]
\[
c_{\psi}^{k}=\frac{2J^{T}_{\psi}}{\pi} - \sum_{i=k+1}^{N}J_{\psi}^{i} + \frac{1}{\pi\sqrt{3}}\sum_{i=k}^{N-1}(\hat{q}_{\phi}^{i})^{2}\Phi_\psi^{i} - \frac{1}{\sqrt{3}}(\hat{q}^{k}_{\phi})^{2} \left(\frac{\Xi^{k}_{\psi}}{\Phi_\psi^{k}} + \frac{\Phi_\psi^{k}}{4\pi}\right) 
\]
\[
c_{\mu}^{k}=\frac{2Q^{T}}{\sqrt{3}\pi} - \frac{4}{\sqrt{3}\pi} \sum_{i=k+1}^{N}Q^{i}  + \frac{1}{2\pi\sqrt{3}}\sum_{i=k}^{N-1}\hat{q}_{\phi}^{i}\Phi_\psi^{i} - \frac{1}{\sqrt{3}} \hat{q}^{k}_{\psi}\left(\frac{\Xi^{k}_{\psi}}{\Phi_\psi^{k}} + \frac{\Phi_\psi^{k}}{4\pi} \right) 
\]
if $\Phi^{k}_{\psi}=0~\wedge~\hat{q}_{\psi}^{k}\ne0$,
\[
c_{\phi}^{k}=\frac{2J^{T}_{\phi}}{\pi} - \sum_{i=k+1}^{N}J_{\phi}^{i} - \frac{1}{3\pi}\sum_{i=k}^{N-1}\hat{q}_{\phi}^{i}\Xi_{\psi}^{i} - \frac{2}{\hat{q}_{\phi}^{k}} \left(\frac{4}{\sqrt{3}\pi}(Q^{k+1}-Q^{k+1}_{M})+\frac{2}{3}\mathcal{Q}^{k+1}_{\psi}-\hat{q}^{k}_{\phi}f^{k-1}(\kappa_{i}) \right)^2
\]
\[
c_{\psi}^{k}=\frac{2J^{T}_{\psi}}{\pi} - \sum_{i=k+1}^{N}J_{\psi}^{i} + \frac{1}{\pi\sqrt{3}}\sum_{i=k}^{N-1}(\hat{q}_{\phi}^{i})^{2}\Phi_\psi^{i} -\hat{q}_{\phi}^{k} \left(\frac{4}{\sqrt{3}\pi}(Q^{k+1}-Q^{k+1}_{M})+\frac{2}{3}\mathcal{Q}^{k+1}_{\psi}-\hat{q}^{k}_{\phi}f^{k-1}(\kappa_{i})\right)
\]
\[
c_{\mu}^{k}=\frac{2Q^{T}}{\sqrt{3}\pi} - \frac{4}{\sqrt{3}\pi} \sum_{i=k+1}^{N}Q^{i}  + \frac{1}{2\pi\sqrt{3}}\sum_{i=k}^{N-1}\hat{q}_{\phi}^{i}\Phi_\psi^{i} - \frac{4}{\sqrt{3}\pi}(Q^{k+1}-Q^{k+1}_{M})+\frac{2}{3}\mathcal{Q}^{k+1}_{\psi}-\hat{q}^{k}_{\phi}f^{k-1}(\kappa_{i})
\]

\emph{(vii)} $\psi$-invariant plane: $\Sigma_{\psi}=\{(r,z)|r=0,\kappa_{2N}<z<+\infty\}$ with rod vector $v=(0,1,0)$.\\
\[
\lambda_{\phi \phi}=\mathcal{O}(1), ~\lambda_{\psi \psi}=\mathcal{O}(r^{2}), ~\lambda_{\phi \psi}=\mathcal{O}(r^{2})
\]
\[
\psi_{\phi}=\mathcal{O}(1), ~\psi_{\psi}=\mathcal{O}(r^{2})
\]
\[
\omega_{a}=\frac{2J^{T}_{a}}{\pi}+\mathcal{O}(r^{2}),~ \mu=\frac{2Q^{T}}{\sqrt{3\pi}} + \mathcal{O}(r^{2})
\]
\[
r\partial_{z}tr(\Psi)=\mathcal{O}(r)
\]
\emph{(viii)} Asymptotic infinity: $\Sigma_{\infty}=\{(r,z)|\sqrt{r^{2}+z^{2}}\to\infty, \frac{z}{\sqrt{r^{2}+z^{2}}}=const\}$\\ \\
The metric fields $\lambda_{ab}$ are given in appendix \ref{ac}, while the remaining ones are given by:
\[
\omega_{a}=\frac{J_{a}^{T}}{\pi}(\frac{\rho^{2}}{r^{2}+z^{2}} - \frac{2z}{\sqrt{r^{2}+z^{2}}}) + \mathcal{O}(\frac{1}{r^{2}+z^{2}})
\]
\[
\mu=\frac{2Q^{T}z}{\pi \sqrt {3} \sqrt{r^{2}+z^{2}}} + \mathcal{O}(\frac{1}{r^{2}+z^{2}})
\]
\[
~\psi_{a}= \mathcal{O}(\frac{1}{\sqrt{r^{2}+z^{2}}})
\]
\[
r\partial_{a}tr(\Psi)dS^{a}= \mathcal{O}(\frac{1}{\sqrt{r^{2}+z^{2}}}).
\] \\ \\
Thus we find that the boundary integral \eqref{E:Mis} vanishes on the rods $I$ and at infinity. Furthermore $\Theta\to0$ at infinity and hence it vanishes everywhere on $\Sigma$, therefore the two field configurations $\Theta_{0}$ and $\Theta_{1}$ coincide with each other. This completes the proof of the uniqueness theorem.\\

\subsection{Application to Exact Solutions with Disconnected Horizons}

In this section we analyze several known solutions of black hole space-times with disconnected horizons. All these solutions satisfy  $\Phi^{k}_{a}=0$, $\mathcal{Q}^{i}_{a}=0$ and $Q^{i}-Q^{i}_{M}=0$. We apply the considerations above to different cases and prove uniqueness of such solutions. We specify the particular rod structure for each case as well as the boundary conditions on the fields $\{\omega_{a}, \psi_{a}\}$.\footnote{The boundary conditions on the metric fields $\lambda_{ab}$ and on the potential $\mu$ are common to all solutions since we do not consider electrically charged solutions in what follows, so we do not write them down explicitly. The same holds for the conditions on any horizon and at infinity.}\\ \\ \\
\textbf{Case 1: Black Saturn with Dipole Ring}
\begin{figure}[H]
\centering
\includegraphics[width=0.5\textwidth]{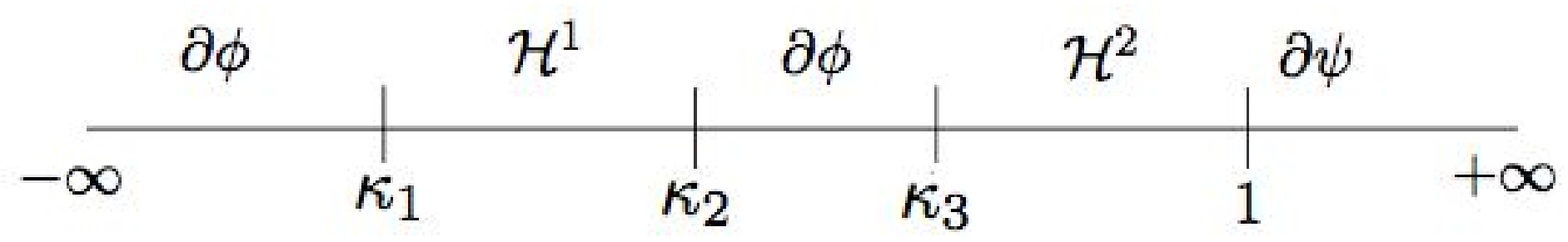}
\end{figure}
This solution was found in \cite{Yazadjiev:2007cd} and it is a dipole charged version of the black saturn found in \cite{Elvang:2007rd}. It describes a Myers-Perry black hole surrounded by a dipole black ring. The regular black saturn solution with dipole charge, after fixing the total mass and angular momenta, exhibits \emph{3-fold continuous non-uniqueness}, this means that one of the rod parameters $\kappa_{i}$ can be expressed in terms of the remaining two. The rod structure and boundary conditions can be summarized as:\\ \\
\emph{(i)}  $\phi$-invariant plane: $\Sigma_{\phi}=\{(r,z)|r=0,-\infty<z<\kappa_{1}\}$ with rod vector $v=(0,0,1)$.\\
\[
\psi_{\phi}=\mathcal{O}(r^{2}),~\psi_{\psi}=\mathcal{O}(1)
\]
\[
\omega_{\phi}=\mathcal{O}(r^{2}),~\omega_{\psi}=-\frac{2J^{T}_{\psi}}{\pi}+\mathcal{O}(r^{2})
\]
\emph{(ii)} $\phi$-invariant plane: $\Sigma_{\phi^{1}}=\{(r,z)|r=0, \kappa_{2}<z<\kappa_{3}\}$ with rod vector $v=(0,0,1)$. \\
\[
\psi_{\phi}=\frac{2q_{\psi}^{1}}{\sqrt{3}} + \mathcal{O}(r^{2}), ~\psi_{\psi}=\mathcal{O}(1)
\]
\[
\omega_{\phi}=\mathcal{O}(r^{2}),~\omega_{\psi}=-\frac{2J^{T}_{\psi}}{\pi} + J^{1}_{\psi} +\mathcal{O}(r^{2})
\]
 \emph{(iii)} BR Horizon: $\Sigma_{\mathcal{H}^{1}}=\{(r,z)|r=0, \kappa_{1}<z<\kappa_{2}\}$ with rod vector $v=(1,0,\Omega^{1}_{\psi})$ where the  $S^{1}$ is parameterized by $\psi$.\\ \\
\emph{(iv)} BH Horizon: $\Sigma_{\mathcal{H}^{2}}=\{(r,z)|r=0, \kappa_{3}<z<1\}$ and rod vector $v=(1,0,\Omega^{2}_{\psi})$.\\ \\
\emph{(v)} $\psi$-invariant plane: $\Sigma_{\psi}=\{(r,z)|r=0,1<z<+\infty\}$ with rod vector $v=(0,1,0)$. \\
\[
\psi_{\phi}=\mathcal{O}(1), ~\psi_{\psi}=\mathcal{O}(r^{2})
\]
\[
\omega_{\phi}=\mathcal{O}(1),~\omega_{\psi}=\frac{2J^{T}_{\psi}}{\pi}+\mathcal{O}(r^{2})
\]
\\
Hence, given the parameters $\kappa_{1},\kappa_{2}$ and the charges $M^{T},J^{T}_{\psi},J^{1}_{\psi},q^{1}_{\psi}$, we obtain a unique black saturn solution.\\ \\
\textbf{Case 2: Black Di-Ring with Dipole Charge}
\begin{figure}[H]
\centering
\includegraphics[width=0.6\textwidth, height=0.07\textheight]{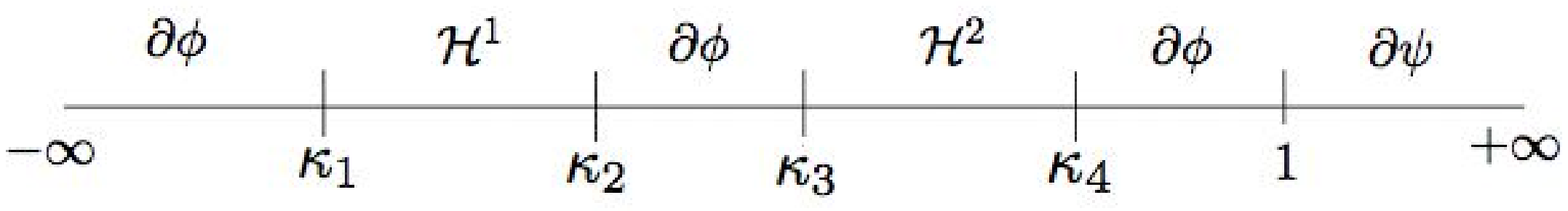}
\end{figure}
This solution describes two concentric black rings with dipole charge \cite{Yazadjiev:2008pt} and it is a dipole charged version of the one found in \cite{Iguchi:2007bd,Evslin:2007fv}. It exhibits \emph{4-fold continuous non-uniqueness} such that there are two independent rod parameters for a regular di-ring. The rod structure and the boundary conditions are summarized as:\\ \\
\emph{(i)}  $\phi$-invariant plane: $\Sigma_{\phi}=\{(r,z)|r=0,-\infty<z<\kappa_{1}\}$ with rod vector $v=(0,0,1)$.\\
\[
\psi_{\phi}=\mathcal{O}(r^{2}), ~\psi_{\psi}=\mathcal{O}(1)
\]
\[
\omega_{\phi}=\mathcal{O}(r^{2}),~\omega_{\psi}=-\frac{2J^{T}_{\psi}}{\pi}+\mathcal{O}(r^{2})
\]
\emph{(ii)} $\phi$-invariant plane: $\Sigma_{\phi^{1}}=\{(r,z)|r=0, \kappa_{2}<z<\kappa_{3}\}$ with rod vector $v=(0,0,1)$. \\
\[
\psi_{\phi}=\frac{2q_{\psi}^{1}}{\sqrt{3}} + \mathcal{O}(r^{2}), ~\psi_{\psi}=\mathcal{O}(1)
\]
\[
\omega_{\phi}=\mathcal{O}(r^{2}),~\omega_{\psi}=-\frac{2J^{T}_{\psi}}{\pi} + J^{1}_{\psi} +\mathcal{O}(r^{2})
\]
\emph{(iii)} $\phi$-invariant plane: $\Sigma_{\phi^{2}}=\{(r,z)|r=0, \kappa_{4}<z<1\}$ with rod vector $v=(0,0,1)$. \\
\[
\psi_{\phi}=\frac{2(q^{1}_{\psi}+ q^{2}_{\psi})}{\sqrt{3}} + \mathcal{O}(r^{2}), ~\psi_{\psi}=\mathcal{O}(1)
\]
\[
\omega_{\phi}=\mathcal{O}(r^{2}),~\omega_{\psi}=\frac{2J^{T}_{\psi}}{\pi} +\mathcal{O}(r^{2})
\]
 \emph{(iv)} BR Horizon: $\Sigma_{\mathcal{H}^{i}}=\{(r,z)|r=0, \kappa_{i}<z<\kappa_{i+1}, ~i=1,3\}$ with rod vector $v=(1,0,\Omega^{i}_{\psi})$ where the  $S^{1}$ is parameterized by $\psi$.\\ \\
\emph{(v)} $\psi$-invariant plane: $\Sigma_{\psi}=\{(r,z)|r=0,1<z<+\infty\}$ with rod vector $v=(0,1,0)$. \\
\[
\psi_{\phi}=\mathcal{O}(1), ~\psi_{\psi}=\mathcal{O}(r^{2})
\]
\[
\omega_{\phi}=\mathcal{O}(1),~\omega_{\psi}=\frac{2J^{T}_{\psi}}{\pi}+\mathcal{O}(r^{2})
\]
 \\
Thus, this solution is uniquely specified given the parameters $\kappa_{1},\kappa_{2}$ and the charges $M^{T},J^{T}_{\psi},J^{1}_{\psi},q^{1}_{\psi},q^{2}_{\psi}$.\\ \\
\textbf{Case 3: Black Bi-Ring}
\begin{figure}[H]
\centering
\includegraphics[width=0.6\textwidth, height=0.07\textheight]{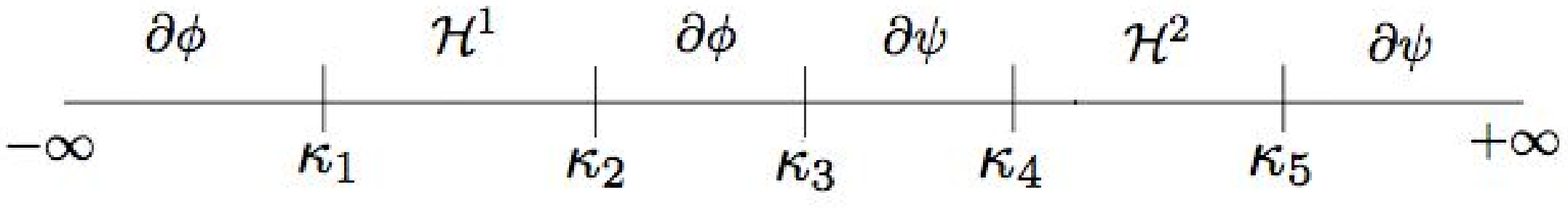}
\end{figure}
This solution describes two black rings placed in different orthogonal planes and was first found in \cite{Elvang:2007hs}. The regular bicycling solution, after fixing the total mass and both angular momenta exhibits \emph{1-fold continuous non-uniqueness}. The rod structure and boundary conditions can be summarized as:\\ \\
\emph{(i)}  $\phi$-invariant plane: $\Sigma_{\phi}=\{(r,z)|r=0,-\infty<z<\kappa_{1}\}$ with rod vector $v=(0,0,1)$.\\
\[
\psi_{\phi}=\mathcal{O}(r^{2}), ~\psi_{\psi}=\mathcal{O}(1)
\]
\[
\omega_{\phi}=-\frac{2J^{T}_{\phi}}{\pi}+\mathcal{O}(r^{2}),~\omega_{\psi}=-\frac{2J^{T}_{\psi}}{\pi}+\mathcal{O}(r^{2})
\]
\emph{(ii)} $\phi$-invariant plane: $\Sigma_{\phi^{1}}=\{(r,z)|r=0, \kappa_{2}<z<\kappa_{3}\}$ with rod vector $v=(0,0,1)$. \\
\[
\psi_{\phi}=\mathcal{O}(r^{2}), ~\psi_{\psi}=\mathcal{O}(1)
\]
\[
\omega_{\phi}=-\frac{2J^{T}_{\phi}}{\pi} + J^{1}_{\phi} +\mathcal{O}(r^{2}),~\omega_{\psi}=-\frac{2J^{T}_{\psi}}{\pi} + J^{1}_{\psi} +\mathcal{O}(r^{2})
\]
\emph{(iii)} BR Horizon: $\Sigma_{\mathcal{H}^{1}}=\{(r,z)|r=0, \kappa_{1}<z<\kappa_{2}\}$ with rod vector $v=(1,\Omega^{1}_{\phi},\Omega^{1}_{\psi})$ where the  $S^{1}$ is parameterized by $\psi$.\\ \\
\emph{(iv)} BR Horizon: $\Sigma_{\mathcal{H}^{2}}=\{(r,z)|r=0, \kappa_{4}<z<\kappa_{5}\}$ with rod vector $v=(1,\Omega^{2}_{\phi},\Omega^{2}_{\psi})$ where the  $S^{1}$ is parameterized by $\phi$.\\ \\
\emph{(v)} $\psi$-invariant plane: $\Sigma_{\psi^{1}}=\{(r,z)|r=0,\kappa_{3}<z<\kappa_{4}\}$ with rod vector $v=(0,1,0)$. \\
\[
\psi_{\phi}=\mathcal{O}(1), ~\psi_{\psi}=\mathcal{O}(r^{2})
\]
\[
\omega_{\phi}=-\frac{2J^{T}_{\phi}}{\pi} + J^{1}_{\phi} +\mathcal{O}(r^{2}),~\omega_{\psi}=-\frac{2J^{T}_{\psi}}{\pi} + J^{1}_{\psi} +\mathcal{O}(r^{2})
\]
\\
\emph{(vi)} $\psi$-invariant plane: $\Sigma_{\psi}=\{(r,z)|r=0,\kappa_{5}<z<+\infty\}$ with rod vector $v=(0,1,0)$. \\
\[
\psi_{\phi}=\mathcal{O}(1), ~\psi_{\psi}=\mathcal{O}(r^{2})
\]
\[
\omega_{\phi}=\frac{2J^{T}_{\phi}}{\pi}+\mathcal{O}(r^{2}),~\omega_{\psi}=\frac{2J^{T}_{\psi}}{\pi}+\mathcal{O}(r^{2})
\]
 \\
Therefore, this solution is unique given the parameters $\kappa_{1}$ and the charges $M^{T},J^{T}_{\phi},J^{T}_{\psi},J^{1}_{\phi},J^{1}_{\psi}$.\\

\subsection{Generalization to Lens Spaces} \label{Lens}
Here we show how this theorem can be generalized to include black hole space-times with Lens space horizon topology. A black hole of this kind has the following rod structure \cite{Jarah:2008bl}:
\begin{figure}[H]
\centering
\includegraphics[width=0.5\textwidth]{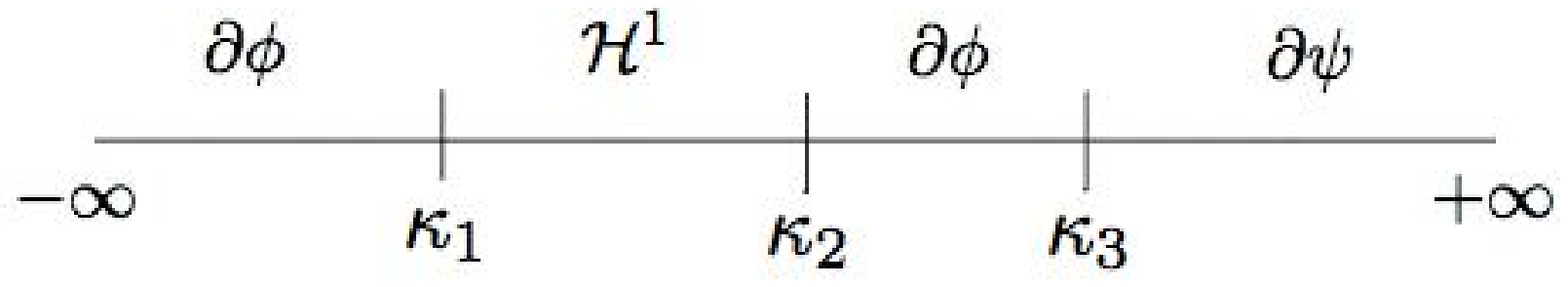}
\end{figure}
The difference between this space-time and a single dipole black ring is on the $\phi$-fixed plane at the right of the horizon rod:\\
\emph{(ii)} $\phi$-invariant plane: $\Sigma_{\phi^{1}}=\{(r,z)|r=0, \kappa_{2}<z<\kappa_{3}\}$ and rod vector $v=(0,1,p)$. \\
As it has been shown in \cite{Tomizawa:2009tb} for this rod we can write:
\begin{equation} \label{E:lens1}
\psi_{\phi}=c_{0} - ph(z) + \mathcal{O}(r^{2}),~\psi_{\psi}=h(z)+\mathcal{O}(r^{2})
\end{equation}
and hence find:
\begin{equation} \label{E:lens2}
\mu=-c_{0} + \frac{2Q}{3\pi} + \mathcal{O}(r^{2})
\end{equation}
\begin{equation} \label{E:lens3}
\omega_{\phi}=-2c_{0}^{2} + pc_{0}h(z)^{2} -\frac{2J_{\phi}}{\pi} + \mathcal{O}(r^{2}),~\omega_{\psi}=-c_{0}h(z)^{2} -\frac{2J_{\psi}}{\pi} + \mathcal{O}(r^{2})
\end{equation}
The problem then relies on finding the constant $c_{0}$. In a similar fashion as in \cite{Yazdjiev:2009kbb} we define at the rightmost semi-infinite rod the magnetic flux:
\begin{equation} \label{E:lens4}
\Phi^{+}=\int_{C^{+}}F=-2\pi\sqrt{3}\psi_{\phi}(\kappa_{4})
\end{equation}
since $\psi_{\phi}$ vanishes as $z\to+\infty$.\\
Hence, using equation \eqref{E:lens1} we can determine the constant $c_{0}$ to be:
\begin{equation} \label{E:lens5}
c_{0}=\frac{1}{2\pi\sqrt{3}}\Phi^{+}
\end{equation}
This then leads to $r\partial_{z}Tr(\Psi)=\mathcal{O}(r)$, therefore, a black hole space-time with Lens space horizon topology is uniquely characterized by its mass, angular momenta, electric charge and magnetic flux.\\ \\
\section{Discussion} \label{discussion}

In this paper we have proved that, in 5D EMCS theory, a non-extremal
asymptotically flat stationary rotating charged black hole solution
with multiple disconnected horizons which, (1) besides the
stationary Killing vector field, admits two mutually commuting axial
Killing vector fields, (2) the topology of each horizon is either
$S^{3}$ or $S^{1}\times S^{2}$, then the solution is
uniquely specified by its rod structure, asymptotic charges,
intrinsic charges and fluxes.
This theorem is a generalization of the theorems given in \cite{Hollands:2007aj,Hollands:2007qf,Hollands:2008fm,Tomizawa:2009ua,Tomizawa:2009tb}, to black hole space-times with disconnected horizons. We have restricted this theorem to include only asymptotic flat solutions, however, this theorem would be trivially generalized to include asymptotic Kaluza-Klein spaces by just defining the necessary fluxes in the leftmost and rightmost semi-infinite rods.\\
In this work we have directed our attention to non-extremal black hole solutions, to include these cases, as it has been shown in
\cite{Figueras:2009ci} for the pure gravity case, a further
specification, namely the near-horizon geometry, has also to be
given in order to define correct boundary conditions. However, if
the near-horizon geometry of all extremal black hole solutions in 5D
EMCS could be fully characterized then our theorem could possibly be
easily generalized to include the extremal case. In fact recent
research on this subject \cite{Kunduri:2009ud} has shed some light
onto this problem but still some more work needs to be done. These
issues deserve further study.

\section*{Acknowledgments}
Jay would like to thanks the Christiania community for providing Denmark with an enriching environment of dazzling creativity. Jay would also like to thanks FCT Portugal for virtually supporting his work over the past six months through the scholarship SFRH/BD/45893/2008 and to all the Gods of Hinduism, Judaism and Islam for not letting him die out of starvation over this same period. We are also very grateful for the useful discussions and nice Hemingway's cocktails with Shinji Hirano, Joan Camps and Simon Ross.


\appendix

\section{Relations between parameterizations of $G_{ij}$} \label{aa}

Here we present the relations between \eqref{E:dsred} and \eqref{E:dswp}.

\[
G_{ij}dx^{i}dx^{j}=-\tau^{-1}r^{2}dt^{2} + \lambda_{ab}(dx^{a} + a^{a}_{t}dt)(dx^{b} + a^{b}_{t}dt)
\]
\[
e^{2\nu}=\tau^{-1}e^{2\sigma}
\]
Therefore we can read off:

\[
G_{tt}=-\tau^{-1}r^{2} + \lambda_{\phi \phi} a^{\phi}_{t} a^{\phi}_{t} +  \lambda_{\psi \psi} a^{\psi}_{t} a^{\psi}_{t} + 2\lambda_{\phi \psi} a^{\phi}_{t} a^{\psi}_{t}
\]
\[
G_{t\phi}=\lambda_{\phi \phi} a^{\phi}_{t}  +  \lambda_{\phi \psi} a^{\psi}_{t}
\]
\[
G_{t\psi}=\lambda_{\psi \psi} a^{\psi}_{t}  +  \lambda_{\phi \psi} a^{\phi}_{t}
\]
\[
G_{\phi \phi}=\lambda_{\phi \phi}
\]
\[
G_{\psi \psi}=\lambda_{\psi \psi}
\]
\[
G_{\phi \psi}=\lambda_{\phi \psi}
\]
The inverse relations are also easily obtained:

\[
\tau=-det(\lambda_{ab})=G_{\phi \psi}^{2}-G_{\phi \phi}G_{\psi \psi}
\]
\[
a^{\phi}_{t}=\tau^{-1}(G_{t \psi}G_{\phi \psi} - G_{t \phi} G_{\psi\psi})
\]
\[
a^{\psi}_{t}=\tau^{-1}(G_{t \phi}G_{\phi \psi}-G_{t \psi} G_{\phi\phi})
\]
The remaining metric coefficients are trivially constructed from the above ones.


\section{Components of $\Psi$} \label{ab}

Here we present the components of the matrix $\Psi$:

\[ \hat{A} = \left( \begin{array}{ccc}
[(1-y)\lambda + (2+x)\psi\psi^{T} - \tau^{-1}\tilde{\omega}\tilde{\omega}^{T} + \mu(\psi\psi^{T}\lambda^{-1}\hat{J} - \hat{J}\lambda^{-1}\psi\psi^{T})] & \tau^{-1}\tilde{\omega} \\
\tau^{-1}\tilde{\omega}^{T} & -\tau^{-1} \end{array} \right)\] \\
\[ \hat{B} = \left( \begin{array}{ccc}
(\psi\psi^{T} - \mu\hat{J})\lambda^{-1} - \tau^{-1}\tilde{\omega}\psi^{T}\hat{J} & [-(1+y)\lambda\hat{J} - (2+x)\mu + \psi^{T}\lambda^{-1}\tilde{\omega})\psi + (z-\mu\hat{J}\lambda^{-1})\tilde{\omega}] \\
\tau^{-1}\psi^{T}\hat{J} & -z \end{array} \right)\] \\
\[ \hat{C} = \left( \begin{array}{ccc}
(1+x)\lambda^{-1} - \lambda^{-1}\psi\psi^{T}\lambda^{-1} & \lambda^{-1}\tilde{\omega}-\hat{J}(z-\mu\hat{J}\lambda^{-1})\psi \\
\tilde{\omega}^{T}\lambda^{-1} + \psi^{T}(z+\mu\lambda^{-1}\hat{J})\hat{J} & [\tilde{\omega}^{T}\lambda^{-1}\omega - 2\mu\psi^{T}\lambda^{-1}\tilde{\omega}-\tau(1+x-2y-xy+z^{2})] \end{array} \right)\] \\
\[ \hat{U} = \left( \begin{array}{ccc}
(1+x-\mu\hat{J}\lambda^{-1})\psi - \mu\tau^{-1}\tilde{\omega} \\
\mu\tau^{-1} \end{array} \right)\] \\
\[ \hat{V} = \left( \begin{array}{ccc}
(\lambda^{-1} + \mu\tau^{-1}\hat{J})\psi \\
\psi^{T}\lambda^{-1}\tilde{\omega} - \mu(1+x-z) \end{array} \right)\] \\
\[ \hat{S}=1+2(x-y) \]
with
\[ \tilde{\omega}=\omega -\mu\psi \]
\[ x=\psi^{T}\lambda^{-1}\psi,~ y=\tau^{-1}\mu^{2}, ~z=y-\tau^{-1}\psi^{T}\hat{J}\tilde{\omega} \]
\[ \hat{J} = \left( \begin{array}{ccc}
0 & 1 \\
-1 & 0 \end{array} \right)\]

\section{Asymptotics} \label{ac}

Here we give the most general asymptotic metric expansion of a 5D black hole space-time in minimal supergravity:

\begin{equation} \label{E:con1}
G_{tt}=-1 + \frac{4M^{T}}{3\pi} \frac{1}{\sqrt{r^{2}+z^{2}}} + \mathcal{O}(\frac{1}{r^{2}+z^{2}})
\end{equation}
\begin{equation} \label{E:con2}
G_{t\phi}=-\frac{J_{\phi}^{T}}{\pi} \frac{\sqrt{r^{2}+z^{2} }-z}{r^{2}+z^{2}} + \mathcal{O}(\frac{1}{r^{2}+z^{2}})
\end{equation}
\begin{equation} \label{E:con3}
G_{t\psi}=-\frac{J_{\psi}^{T}}{\pi} \frac{\sqrt{r^{2}+z^{2} }+z}{r^{2}+z^{2}} + \mathcal{O}(\frac{1}{r^{2}+z^{2}})
\end{equation}
\begin{equation} \label{E:con4}
\lambda_{\phi \phi}=(\sqrt{r^{2}+z^{2}}-z)(1 + \frac{2(M^{T} + \eta)}{3\pi \sqrt{r^{2}+z^{2}}} + \mathcal{O}(\frac{1}{r^{2}+z^{2}}))
\end{equation}
\begin{equation} \label{E:con5}
\lambda_{\psi \psi}=(\sqrt{r^{2}+z^{2}}+z)(1 + \frac{2(M^{T} - \eta)}{3\pi \sqrt{r^{2}+z^{2}}} + \mathcal{O}(\frac{1}{r^{2}+z^{2}}))
\end{equation}
\begin{equation} \label{E:con6}
\lambda_{\phi \psi}=\zeta \frac{r^{2}}{\sqrt{r^{2}+z^{2}}} + \mathcal{O}(\frac{1}{r^{2}+z^{2}})
\end{equation}



\providecommand{\href}[2]{#2}\begingroup\raggedright\endgroup

\end{document}